\documentclass[twocolumn,a4paper,showpacs,prl]{revtex4}
\usepackage{graphicx}
\usepackage{verbatim}
\usepackage{amsmath}
\usepackage{dcolumn}
\usepackage{bm}

\begin{document}
\title{Ridge Network in Crumpled Paper} 

\author{Christian Andr\'{e}  Andresen}
\email{Christian.Andresen@ntnu.no}
\author{Alex Hansen}
\email{Alex.Hansen@ntnu.no}
\affiliation{Department of Physics, Norwegian University of Science and Technology, N--7491 Trondheim, Norway}
\author{Jean Schmittbuhl}
\email{Jean.Schmittbuhl@eost.u-strasbg.fr}
\affiliation{Institut de Physique du Globe Strasbourg, 5 rue Ren\'{e} Descartes, 67084 Strasbourg, France}


%83.60.-a :    Material behavior
%89.75.Hc :    Networks and genealogical trees
%89.75.Fb :    Structures and organization in complex systems
%42.62.-b :    Laser applications

\begin{abstract}
The network formed by ridges in a straightened sheet of crumpled paper is studied using a laser profilometer. Square sheets of paper were crumpled into balls, unfolded and their height profile measured. From these profiles the imposed ridges were extracted as networks. Nodes were defined as intersections between ridges, and links as the various ridges connecting the nodes. Many network and spatial properties have been investigated. The tail of the ridge length distribution was found to follow a power-law whereas the shorter ridges followed a log-normal distribution. The degree distribution was found to have an exponentially decaying tail, and the degree correlation was found to be disassortative. The facets created by the ridges and the Voronoi diagram formed by the nodes have also been investigated.

\end{abstract}

\maketitle

\section{Introduction}

The crumpling of paper is an everyday occurrence, yet it is a surprisingly rich and complex process. Paper is an elastic, flexible and heterogeneous material and many authors have tried to describe its crumpling properties analytically \cite{art:Amar,art:Lobkovsky1997,art:Wood}, numerically \cite{art:Vliegenthart,art:Plouraboue} and experimentally \cite{art:Blair,art:Houle,art:Sultan,art:Boue,art:sethna,art:Balankin}. The crumpling process of paper is also interesting because it is a special case of the thin plate deformation problem that is central in describing processes that occur for example in car crashes and tank failures \cite{art:Lobkovsky1995}. Earlier studies have tried to describe the ridge network of crumpled paper \cite{art:Blair,art:Plouraboue,art:Balankin} and some results have been found, however much is still unclear. This work aims at describing the ridge network formed during a common hand crumpling process of ordinary printing paper. The application of modern network theory \cite{art:Boccaletti,art:Newman,art:Sneppen} have been specially emphasized.

This paper is organized as follows. In section 2 the experimental procedure is described, and in section 3 the ridge detection method is presented. The results are discussed in section 4, in particular the ridge length and the degree distribution are discussed. Also the degree-degree correlation, the clustering and the surface roughness is investigated in addition to the facet distribution and the angular ridge distribution. Finally the main conclusions are summarized in section 5. 

\section{Experimental procedure}

Ordinary printing paper was used for all the experiments, and some of the properties of the paper are given in table \ref{tab:samples}. All the samples were cut into square sheets of 21 x 21 cm, and crumpled by hand into small balls. The diameter of the various balls produced are given in table \ref{tab:samples}. The hand crumpling procedure have been applied before \cite{art:Blair,art:Houle,art:Balankin}, and is practical because it is easy to conduct and produces a compact result. Unfortunately the process in not repeatable and poorly controlled. Several test crumplings were conducted before the measured samples were crumpled in order to reduce the variance between the samples. Earlier studies \cite{art:Houle} on acoustic emissions from crumpling of various materials have indicated that the emission spectra show a surprisingly low sensitivity to the crumpling method. This may indicate that the outcome of the crumpling is not highly sensitive to the details of the process. Balankin \textit{et al.} \cite{art:Balankin} discuss the scaling behavior of the crumpling process for different paper thicknesses. They conclude that the impact of the variation of the applied confinement force $F$ on the ball radius $R$ is small since there is only a weak dependence $R \propto F^{-0.25}$. For these reasons no special precautions, such as dents or initial folding, were taken to increase repeatability. After crumpling the samples, they were carefully unfolded taking care not to tear the paper, introduce new ridges or remove some of the original ridges. When the paper ball was unfolded the paper was stretched to a size of 20 x 20 cm, and fastened to an aluminum plate. This ensured that the vertical height of the samples were no more than 12 mm (the maximum range for the instrument used). 

\begin{figure}
\centering
\begin{tabular}{c}
\begin{minipage}{0.95\columnwidth}
\centering
\includegraphics[width=\columnwidth, angle=0]{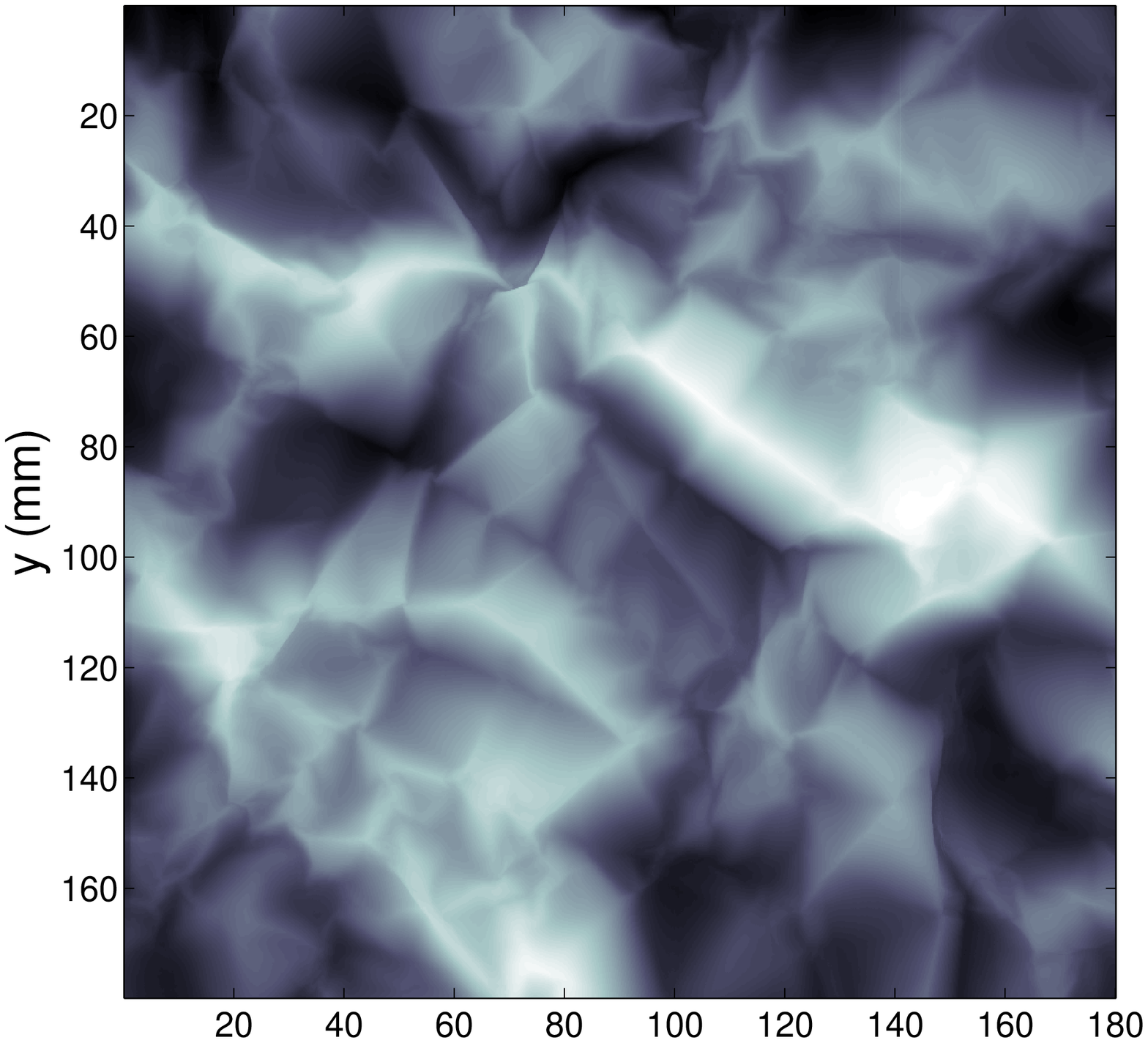}
\end{minipage}
\\
\begin{minipage}{0.95\columnwidth}
\centering
\includegraphics[width=\columnwidth ]{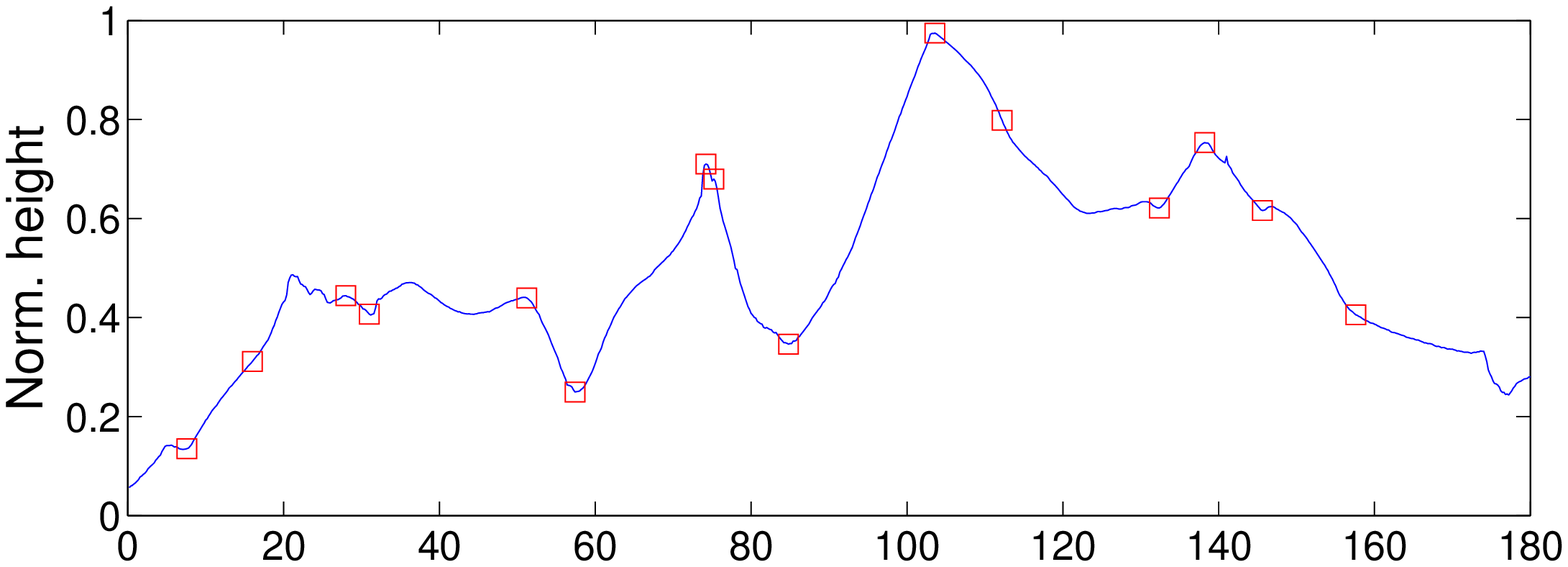}
\end{minipage}
\\
\begin{minipage}{0.95\columnwidth}
\centering
\includegraphics[width=\columnwidth ]{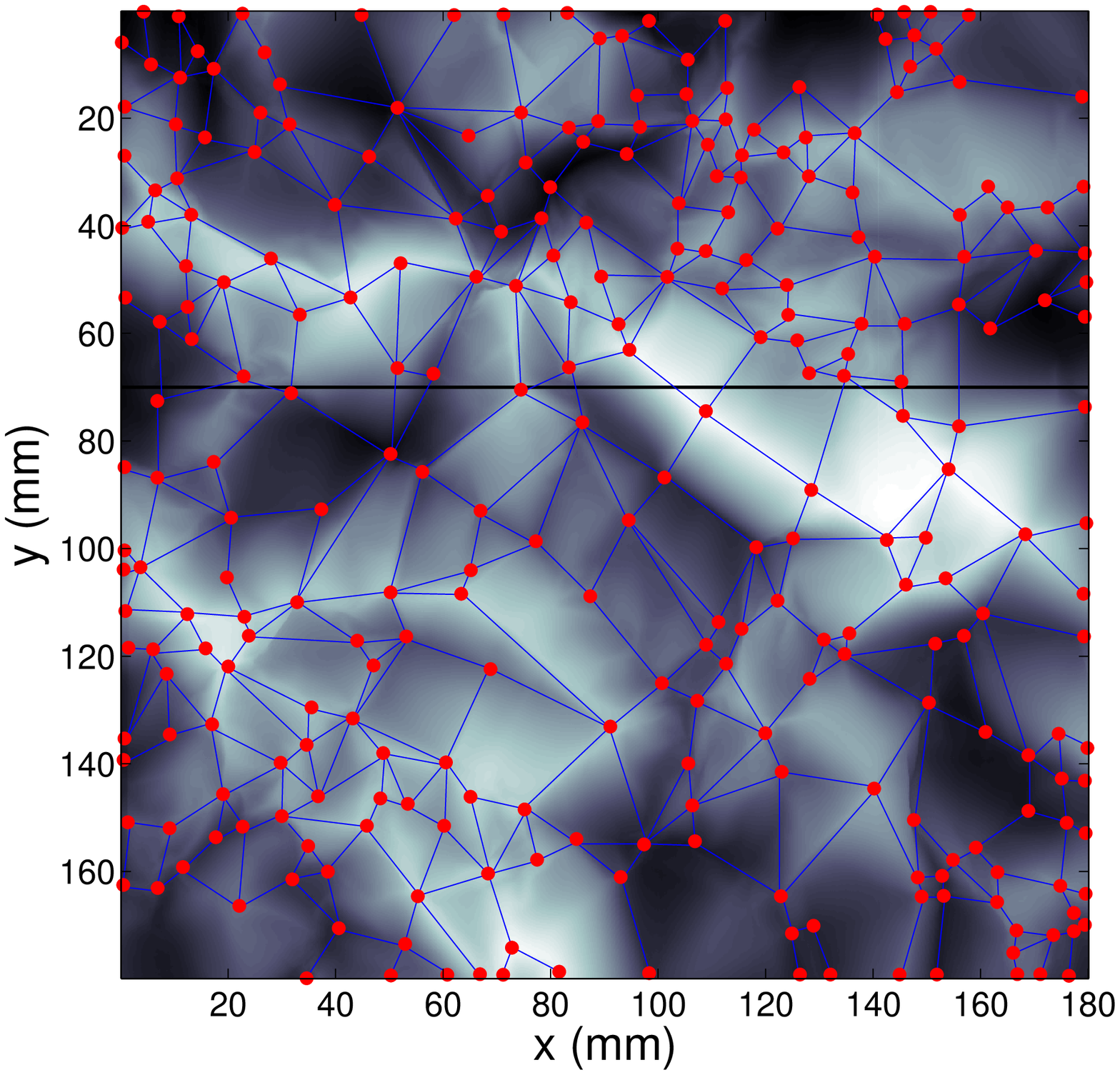}
\end{minipage}
\end{tabular}
\caption{(Color On-line) Top: Grey scale height plot of sample 6 as function of x and y position. Color indicate elevation (brighter is higher), ridges are clearly visible. Middle: A single one dimensional profile from sample 6 (marked as a black line in the bottom plot). The points of the profile that give rise to ridges are marked by squares. Bottom: Network extracted from sample 6 superimposed on the gray scale plot from the top figure.}
\label{fig:map}
\end{figure}

The full (2+1)-dimensional height mapping was measured profile by profile using a laser profilometer over an area of 18 x 18 cm in the center of the samples. The height of each point was measured using a laser giving a voltage output linearly proportional to the distance between the probe and the paper surface. The voltage output was converted to a floating point length-measure using a 16 bit AD converter. The laser diameter used was 30 $\mu$m, however accuracy considerably smaller than this could be achieved. Each profile was acquired by sliding the sample under the probe while measuring. Multiple profiles were acquired by stepping the probe normal to the sliding direction. A typical one-dimensional height profile and a complete (2+1)-dimensional map are shown in figure \ref{fig:map}. The number of points per profile was kept equal to the number of profiles resulting in a square grid of measurements or "pixels". The number of points used for the various samples are given in table \ref{tab:samples}. The in-plane accuracy of each point was no larger than 10 $\mu$m for any sample, and in the out of plane direction it was 0.5 $\mu$m for all samples. 

\begin{table}
\begin{center}
\begin{tabular}[c]{c c c r @{ $\pm$ } l r @{ $\pm$ } l c}
Sample & X step   & Y step   &  \multicolumn{2}{c}{Thickness} & \multicolumn{2}{c}{Weight} &  Ball diam. \\
       &          &          &  \multicolumn{2}{c}{[ $\mu$m]}     & \multicolumn{2}{c}{[g/m$^2$]}    &   [mm]        \\
\hline
1      &  900     &   900    &   51 & 5                            & 49.0  & 1.0                      &   26 $\pm$ 2  \\
2      & 1800     &  1800    &   51 & 5                            & 50.0  & 1.0                      &   27 $\pm$ 2  \\
3      & 1000     &  1000    &   95 & 5                            & 80.0  & 0.5                      &   35 $\pm$ 2  \\
4      &  900     &   900    &   95 & 5                            & 80.0  & 0.5                      &   32 $\pm$ 2  \\
5      & 3600     &  3600    &  100 & 2                            & 83.0  & 0.5                      &   33 $\pm$ 2  \\
6      &  900     &   900    &  220 & 5                            & 175.0 & 1.0                      &   43 $\pm$ 2  \\

\end{tabular}
\caption{List of samples investigated. X steps and Y steps is the number of measured points in the X and Y direction respectively. Thickness is the thickness of the paper and ball diameter is the diameter of the ball produced during the crumpling process. All samples was originally 21 x 21 cm and thereafter unfolded and stretched to 20 x 20 cm producing a maximum height of 12 mm. An area of 18 x 18 cm in the center of the samples were measured.}
\label{tab:samples}
\end{center}
\end{table}

\begin{table}
\begin{center}
\begin{tabular}[c]{c c c c c c c}
Sample & Nodes & Links   & C            &      $C_D$    &     $C_R$      &   Max Deg. \\
\hline
1      & 503   &   890   &   0.182      &     0.4371         &  0.0045    &   8         \\
2      & 1211  &   2238  &   0.190      &     0.4315         &  0.0020    &   9         \\
3      & 190   &   293   &   0.138      &     0.4458         &  0.0095    &   6         \\
4      & 350   &   580   &   0.162      &     0.4394         &  0.0064    &   8         \\
5      & 929   &   1829  &   0.231      &     0.4326         &  0.0029    &   10        \\
6      & 286   &   501   &   0.199      &     0.4384         &  0.0083    &   8         \\

\end{tabular}
\caption{List of extracted networks with their number of nodes, number of links, clustering coefficient, C, the clustering coefficient for the corresponding planar Delaunay network, $C_D$, the clustering coefficient for a non-planar randomized network with the same degree distribution, $C_R$, and the maximum node degree for the network, Max Deg. The clustering coefficients for the random networks was calculated using an average over 1000 samples after each sample had 10000 random rewiring.}
\label{tab:nets}
\end{center}
\end{table}

\section{Ridge detection}

A ridge stands out as a line of high curvature in a otherwise smooth landscape. The curvature of any point in a height profile $\xi ( \vec x)$ can be calculated as the field $\bigtriangledown^2 \xi(\vec x)$, where $\vec x = (x,y)$ is the planar coordinates. In the present case it proved necessary to smooth the height profile $ \xi ( \vec x)$ with a short range gaussian filter before calculating $\bigtriangledown^2 \xi(\vec x)$ in order to filter away small scale features. A range of different filters were tested, and the result did not seem sensitive to the details of the filter. The main effect of the filtering was the removal of single isolated high-curvature pixels, or small groups of such, and a narrowing of the ridge lines. After filtering, the curvature field was calculated and thresholded so that all points over a given value were considered to have a unit value and all other points to have a zero value. Any isolated points above the threshold were filtered away. From the remaining points lines were detected as ridges. It is throughout this paper assumed that all ridges are straight lines. It proved difficult to automate the ridge extraction process from the thresholded field, finally this step had to be done manually. Some statistics of the produced networks are listed in table \ref{tab:nets}. Figure \ref{fig:map} shows an example of a full ridge network. In the middle plot of figure \ref{fig:map} a single one-dimensional profile is given, and all points along this profile giving rise to ridges are marked. It can be noted from this figure that not all sections of the profile that have high curvature give rise to a ridge, while some smooth sections does give rise to a ridge. This may stem from the directionality of the ridges relative to the profile shown. Ridges crossing the profile at a small angle may seem smooth, but small local dents crossing close to orthogonally may seem large. 

Nodes are defined as intersections between ridges, and a ridge therefore only extends from one node to another. All the links are regarded as undirected since a paper ridge does not have any preferred direction. The networks formed are fully connected and have therefore only one component. 

\section{Network properties}

The different paper thicknesses used in the experiments showed a clear trend that thinner paper crumple more than thick paper, and therefore produce more nodes and links (see table \ref{tab:samples} and \ref{tab:nets}). Apart from the scale of the network created, no significant differences in the various distributions referred to below was detectable. As a consequence most distributions are averaged over all samples after each of them have been normalized appropriately. The lack of change in the behavior due to sample thickness may arise from the small amount of data available, and no correspondence between paper thickness and other properties can be excluded. From a scaling point of view a qualitative change of behavior is not expected since a large and thick sheet of paper is equivalent to a thin and small sheet. Note that the needed confinement also varies with the paper thickness, and all our experiments are conducted at approximately the same confinement. 
Sultan and Boudaoud \cite{art:Sultan} discuss two regimes for the crumpling process depending on the confinement of the sample. The transition confinement is partly dependent on the paper thickness. Our experiments are as mentioned conducted at approximately constant confinement (although it is poorly controlled), and it might therefore be that due to the varying paper thickness our samples lie in different regimes. However the uniform behavior of the samples indicates that they are all in the same regime. Also the number of self-contacts are very large for all the samples, and this indicates that they are all in the highly confined regime. 

It can be seen from table \ref{tab:samples} that samples 3 and 4 both have the same paper thickness, although they have a significantly different number of links and nodes. This is most likely due to the difference in confinement. Sample 3 had a larger ball radius than sample 4, and was therefore less confined, and have also fewer nodes and links than sample 4.

\subsection{Ridge length}

\begin{figure*}
\centering
%\begin{tabular{l}
%a)\\
\begin{tabular}{l l}
a) & b)\\
\begin{minipage}{\columnwidth}
\centering
\includegraphics[width=0.95\columnwidth, angle=0]{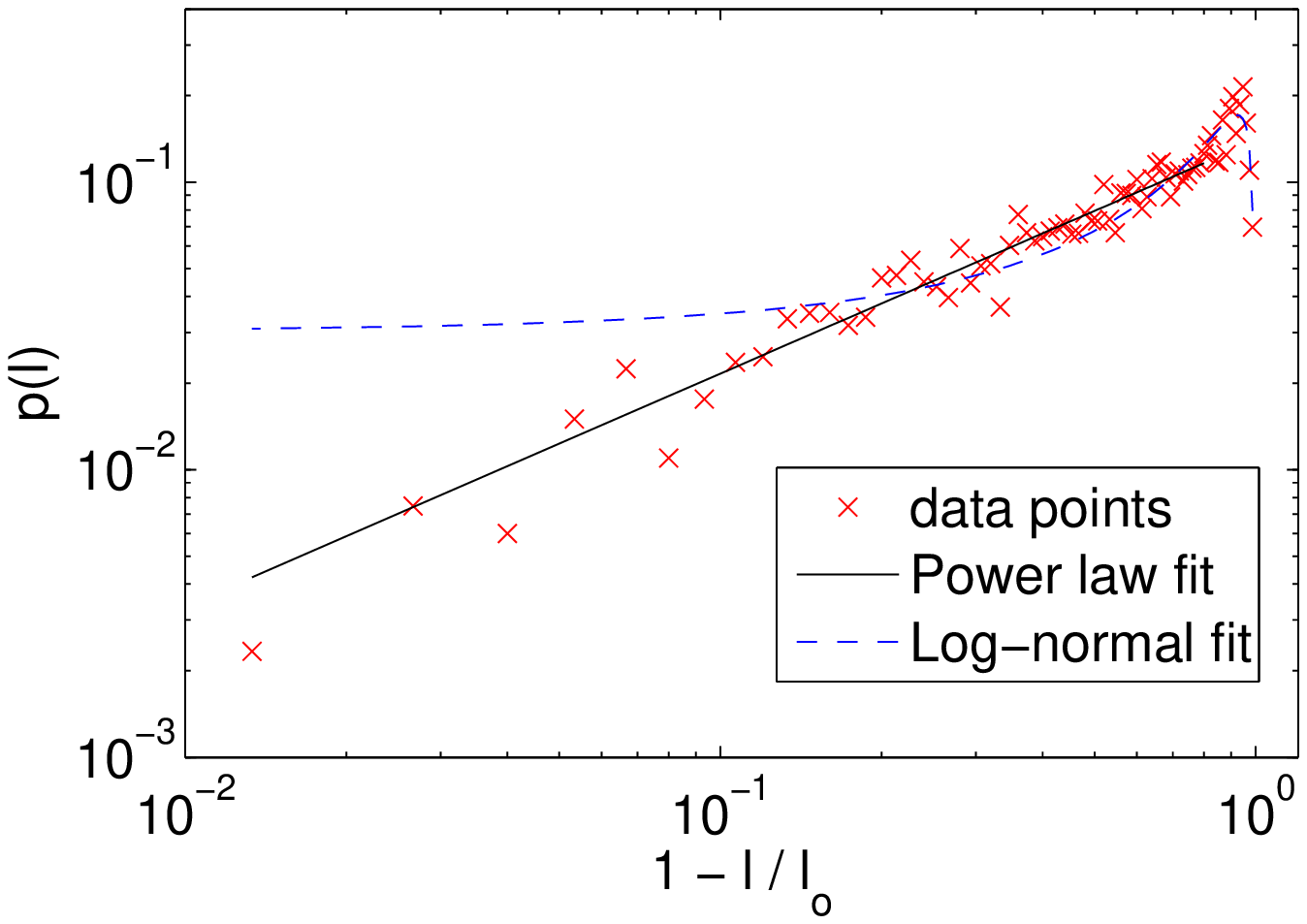}    
\end{minipage}
&
\begin{minipage}{\columnwidth}
\centering
\includegraphics[width=0.95\columnwidth, angle=0]{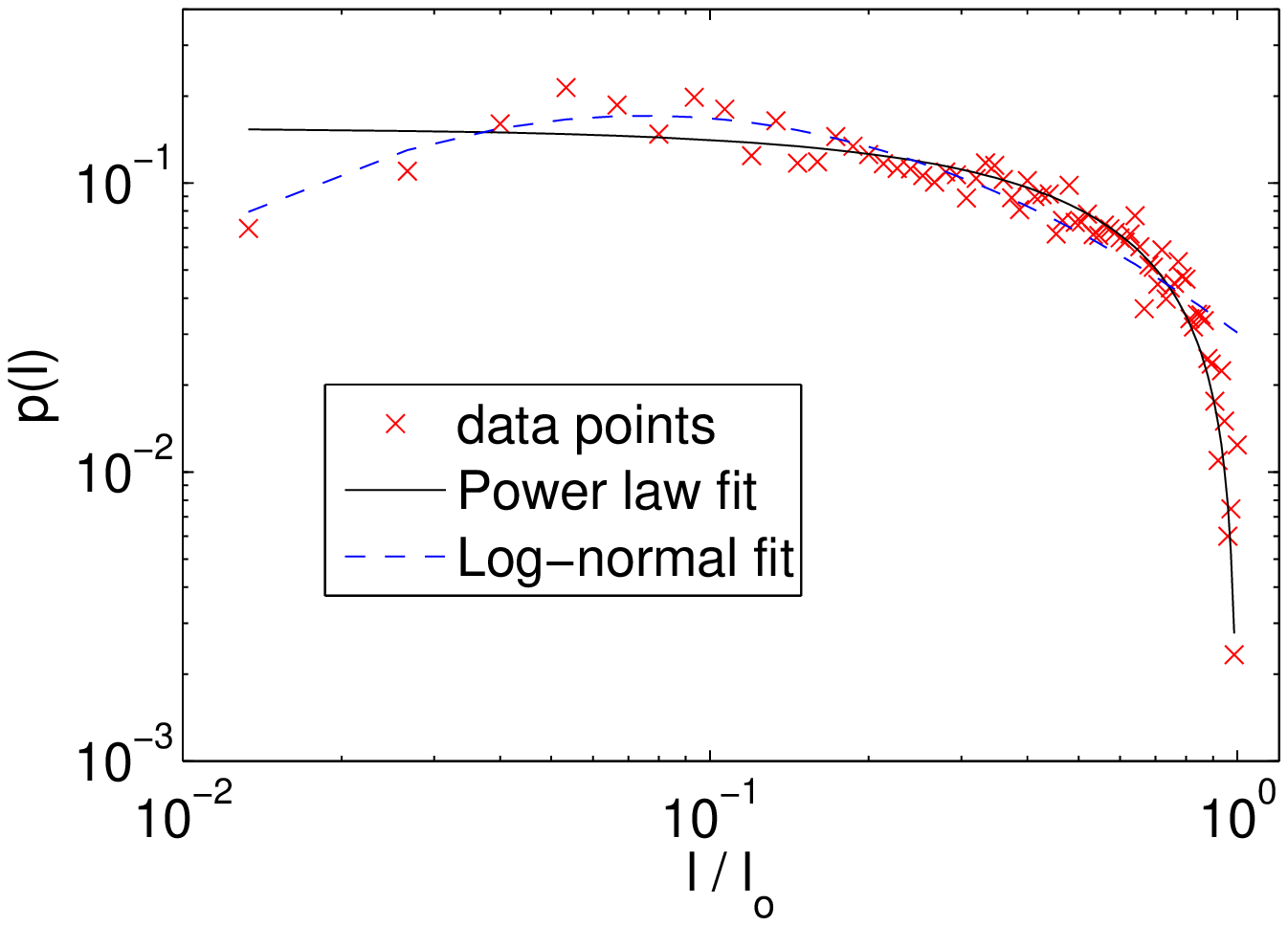}  
\end{minipage}
\end{tabular}
\caption{(Color On-line) a) Plot of the average non-cumulative ridge length distribution, $p(l)$, as a function of $1 - l / l_0$ where $l$ is the ridge length and $l_0$ is the maximum ridge length for any given sample. The data is fitted by a log-normal and a power-law $p(l) \propto (1- l/l_0)^{\beta}$ with $\beta = 0.81$. b) Plot of the average non-cumulative ridge length distribution, $p(l)$, as a function of normalized ridge length $l / l_0$.} 
\label{fig:rlength}
\end{figure*}

\begin{figure}
\centering
\includegraphics[width=0.95\columnwidth, angle=0]{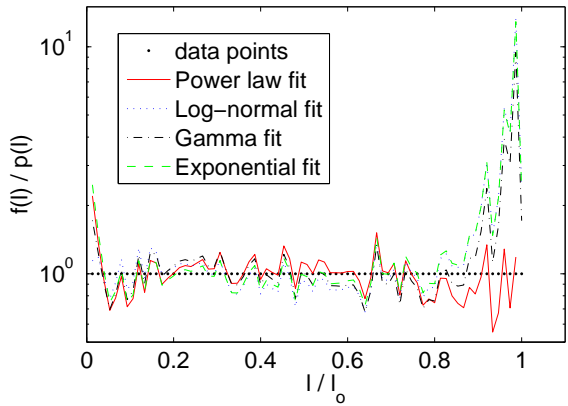} 
\caption{(Color On-line) For comparison the results shown in figure \ref{fig:rlength} from fitting the ridge length distribution, $p(l)$, with a gamma, log-normal and exponential function divided by the data itself are shown together with the same plot for the power-law fit as a function of normalized ridge length $l / l_0$.}
\label{fig:rlength_insert}
\end{figure}

The length of a ridge between node $a$ and $b$ is defined as the spatial length from node $a$ to $b$, following the assumption that all ridges are straight lines. Previous works have reported log-normal, gamma and exponential functions \cite{art:Blair,art:Wood,art:Balankin} to give good fits for this distribution. However we find that, whereas the small scale part of the distribution is well fitted by a log-normal function, the tail of the distribution is not well fitted by any of the above mentioned functions. The large scale part of the distribution is better fitted by a power-law function $p(l) \propto (1 - l/l_0)^{\beta}$ where $l$ is the ridge length and $l_0$ is the maximum ridge length for a given sample. Both fits can be seen in figure \ref{fig:rlength}. We have found the tail to be best fitted by an exponent $\beta = 0.81$. To compare the fits of the different functions they are plotted in figure \ref{fig:rlength_insert} divided by the original distribution in order to emphasize any discrepancies.

The underlying reason for the shift in behavior may stem from the fact that the distribution of short ridges is dominated by remnants of originally long ridges. These ridges have been intersected by "younger" ridges crossing them after their formation. As outlined by Blair and Kudrolli \cite{art:Blair} this random sectioning of ridges will give rise to a log-normal length distribution. The larger ridges on the other hand have not been so heavily sectioned by younger ridges. They are therefore not expected to follow the log-normal distribution of the shorter ridges. Instead we detect a power-law dependency of the distribution of the difference between the longest ridge $l_0$ and the ridge length. It is reasonable to assume that larger samples will produce larger maximum ridges, and therefore $l_0$ is a sample size dependent quantity. Why this difference should exhibit a scale-free behavior is not clear.

\subsection{Degree distribution}

\begin{figure}
\centering
\includegraphics[width=0.95\columnwidth, angle=0]{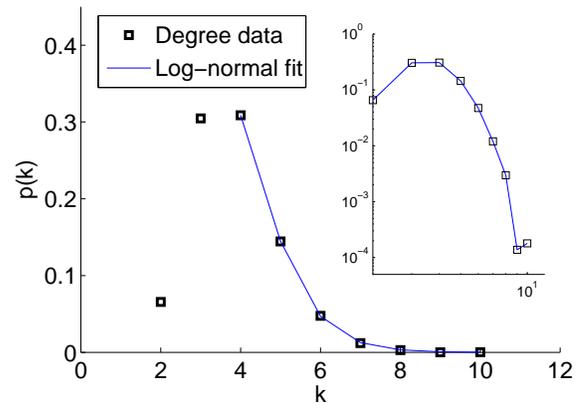}
\caption{(Color On-line) Plot of the degree distribution, $p(k)$, as a function of node degree $k$ with a fitted log-normal tail. The insert shows the same data plotted on log-log scale. This shows that the distribution does not have a power-law distributed tail.}
\label{fig:kdist}
\end{figure}

The degree of a node is defined as the number of ridges meeting at that node. The distribution has been found to have a maximum probability at a median degree and produce a gaussian like form that is plotted in figure \ref{fig:kdist}. The tail of the distribution is well fitted by a log-normal function of the same form as in equation \ref{eq:log_normal}. This is in strong contrast to many naturally occurring networks that show a power-law tail giving a larger portion of high degree nodes than can be seen in the acquired samples.

\subsection{Degree-degree correlation}

The correlation between the degree of connected nodes has been studied using the procedure developed by Maslov and Sneppen \cite{art:Sneppen}. They have defined a correlation measure 

\begin{equation}
C(k_1,k_2) = \frac{P(k_1,k_2)}{P_R(k_1,k_2)},
\end{equation}

where $P(k_1,k_2)$ is the probability that a node of degree $k_1$ is linked to a node of degree $k_2$. $P_R(k_1,k_2)$ is the same average probability for a set of randomized networks. The randomized networks are assumed to have the same number of nodes and links, and the same degree distribution as the original network. A value $C(k_1,k_2) > 1$ indicates that there is an over-representation of links between nodes with degree $k_1$ and $k_2$, whereas $C(k_1,k_2) < 1$ indicates an under-representation. In order to look at the statistical signification of the correlation, Maslov and Sneppen introduced another correlation measure 

\begin{equation}
Z(k_1,k_2) = \frac{P(k_1,k_2) - P_R(k_1,k_2)}{\sigma_R(k_1,k_2)},
\end{equation}

where $\sigma_R(k_1,k_2)$ is the standard deviation of the samples used to generate $P_R(k_1,k_2)$. For $P(k_1,k_2)$ only the sample data is available. If a given coupling $P(k_1,k_2)$ is over-represented (that is $P(k_1,k_2) > P_R(k_1,k_2)$) then $Z(k_1,k_2) > 0$ and if it is under-represented $Z(k_1,k_2) < 0$. If the standard deviation is small the corresponding correlation coefficients are large, thus emphasizing statistically significant results. In all results presented here 1000 randomized versions of the various samples were used to produce $P_R(k_1,k_2)$ and $\sigma_R(k_1,k_2)$. Each randomization used 10000 rewirings of the original network.

\begin{figure}
\centering
\includegraphics[width=\columnwidth]{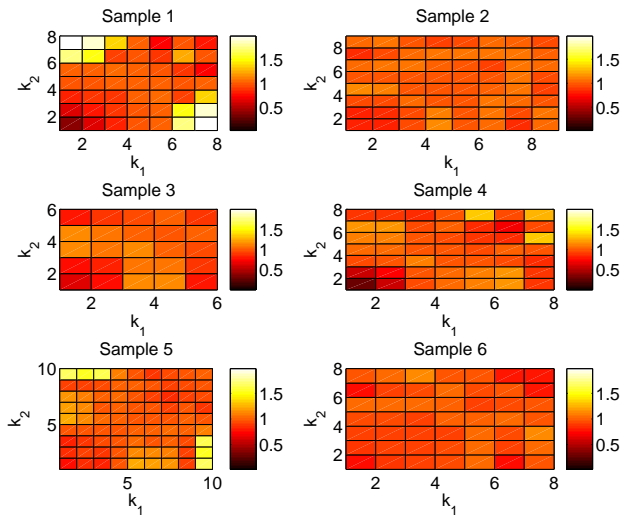}
\caption{(Color On-line) Plot of the correlation matrix $C(k_1,k_2) = P(k_1,k_2) / P_R(k_1,k_2)$ for each sample. The plots indicate that the networks are disassortative.}
\label{fig:degdeg}
\end{figure}

\begin{figure}
\centering
\includegraphics[width=\columnwidth]{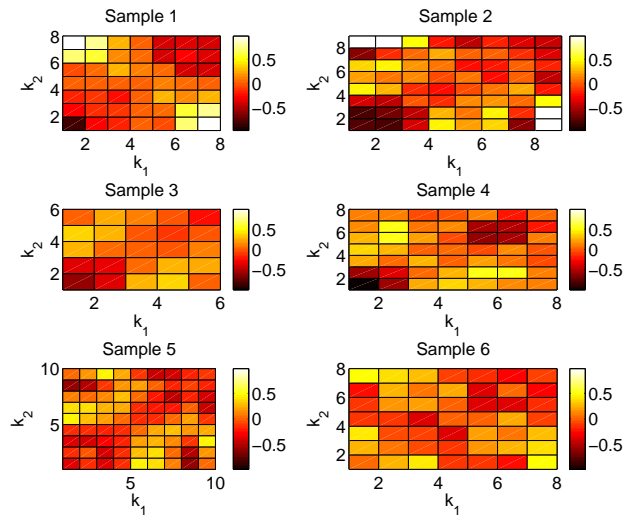}
\caption{(Color On-line) Plot of the Z-matrix $Z(k_1,k_2) = (P(k_1,k_2) - P_R(k_1,k_2))/\sigma_R(k_1,k_2)$ for each sample. As in the $C(k_1,k_2)$ case of figure \ref{fig:degdeg} the plot indicates a disassortative trend.}
\label{fig:degdegZ}
\end{figure}

Figure \ref{fig:degdeg} shows the $C(k_1,k_2)$ matrix for all the samples. There is a tendency of small degree nodes not to link to other small degree nodes, but rather link to large degree nodes. Links between large degree nodes is also under-represented. This type of networks is known as disassortative networks. Figure \ref{fig:degdegZ} shows the $Z(k_1,k_2)$ matrices for the same samples, and the same trends as in figure \ref{fig:degdeg} can be observed. There is a clear trend in nearly all examined networks \cite{art:round_table} that technical and biological networks such as the Internet and various protein interaction networks are disassortative, and that social networks such as acquaintance networks are assortative. The underlying reason for this is still not fully understood.

\subsection{Clustering}

The cluster coefficient $C$ for all the samples are given in table \ref{tab:nets}, and they are all in the range 0.13 to 0.23. The definition used here is the standard 

\begin{subequations}
\begin{equation}
C = \frac{1}{N} \Sigma_{i=1}^{i=N} C_i,
\end{equation}
\begin{equation}
C_i = \frac{2 E_{\textrm{NN}}}{k_i(k_i-1)},
\end{equation}
\end{subequations}

where $N$ is the number of nodes in the network, $E_{\textrm{NN}}$ is the number of links between nearest neighbors of node $i$ and $k_i$ is the degree of node $i$ \cite{art:Newman}. A network embedded in two-dimensional Euclidean space with no crossing links is called a \textit{planar network}, and is described by West \cite{art:West}. Generating a planar randomized network for comparing the cluster coefficients is very hard since no links can cross and the rewiring therefore must be local. However the clustering can be compared with the Delaunay network \cite{art:Aurenhammer} for the same spatial layout of nodes. For a given spatial node configuration and degree distribution the Delaunay network gives the maximum possible clustering coefficient. The clustering coefficient for a Delaunay network made from nodes randomly distributed in the plane and with a number of nodes comparable to our samples is 0.44. Delaunay networks are closely linked to Voronoi diagrams, and both are described bellow. The cluster coefficient for a non-planar random network, where the links can cross, having the same number of nodes and links and the same degree distribution is in the order of 0.001. The ridge networks have a much higher clustering than the non-planar networks. This is expected because any node in a planar network has a low chance of being linked to a far away node. This will generally increase the local clustering \cite{art:West}. On the other hand the clustering is significantly lower than in the Delaunay case. This indicates that the ridge networks does not form highly interconnected cliques. 

\section{Geometrical properties}

Various geometrical properties of crumpled thin sheets have been investigated in the past \cite{art:Blair,art:Cerda,art:Chaieb}. Here we discuss the size distribution of facets formed by the ridges and of the Voronoi sections formed by the location of the nodes. The angular distribution of the ridges and the 3-cone structures are also investigated.

\subsection{Facets}
\begin{figure}
\centering
\includegraphics[width=0.9\columnwidth,angle = 0]{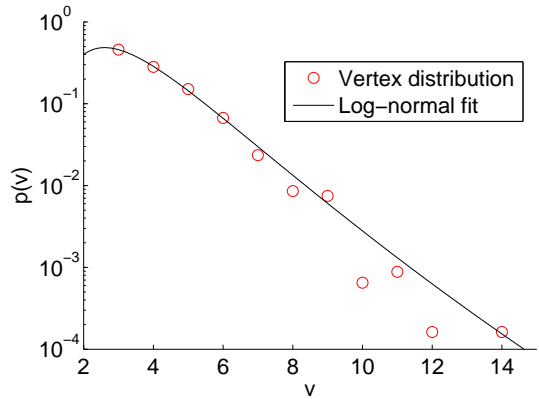}
\caption{(Color On-line) Plot of the average vertex distribution, $p(v)$, as a function of the number of vertexes $v$ for the facets formed by the ridges. The data is fitted to a log-normal function with $\sigma = 0.42$ and $\mu = 1.13$. }
\label{fig:vert}
\end{figure}

The nodes and links of the network form facets (also called domains) of various sizes and shapes. A facet is defined as an area of the crumpled paper confined by a closed loop of ridges that is simply connected, meaning that it contains no internal  facets. The nodes bordering the facets are the corners or vertexes of the facet. The distribution of facet areas and number of vertexes for each sample have been calculated. The vertex distribution for all the samples was averaged giving each sample equal weight. The number of facets with 3, 4, 5 and 6 vertexes was 46\%, 28\%, 15\% and 8\% respectively, and the number of facets with more than 6 vertexes was 4\%. The maximum number of vertexes was 14. In figure \ref{fig:vert} the distribution of the facet vertex number can be seen, the data is fitted with a log-normal function 

\begin{equation}
p(a) = \frac{1}{\sqrt{2 \pi} a \sigma} e^{-(ln(a) - \mu)^2/(2 \sigma)^2},
\label{eq:log_normal}
\end{equation}

where $a$ is the vertex number, $\mu$ is the logarithm of the average number of vertexes per facet and $\sigma$ is the standard deviation. The best fit was achieved with $\sigma = 0.42$ and $\mu = 1.13$. 

\begin{figure}
\centering
\includegraphics[width=0.9\columnwidth, angle = 0]{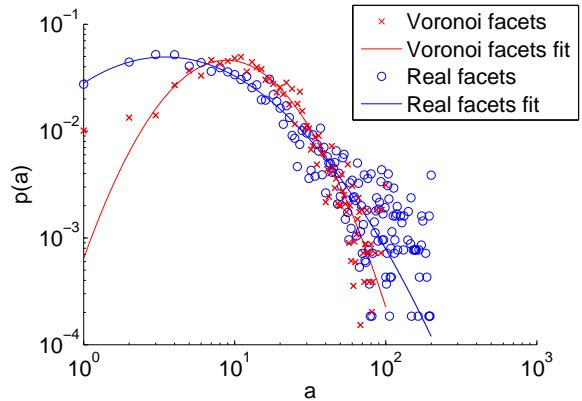}
\caption{(Color On-line) Plot of the average area distribution, $p(a)$, as a function of the facet area $a$ for the facets formed by the ridges and the Voronoi regions. Both data are fitted by a log-normal function with $\sigma = 1.17$ and $\mu = 2.61$ for the facets and $\sigma = 0.74$ and $\mu = 2.73$ for the Voronoi regions, both in the same arbitrary units of area. }
\label{fig:area_facets}
\end{figure}

The areas of the facets have also been investigated. The binned distribution of areas was normalized by the maximum area for each sample, and the average over all samples calculated. The resulting distribution can be seen in figure \ref{fig:area_facets} together with a log-normal fit. The best fit parameters were $\sigma = 1.17$ and $\mu = 2.16$ in arbitrary units.

\subsection{Voronoi networks}
\label{sec:voronoi}
\begin{figure}
\begin{center}
\includegraphics[width=1.1\columnwidth]{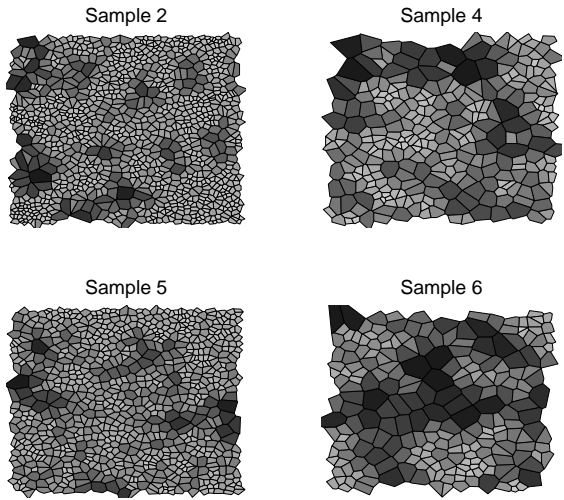}
\caption{(Color On-line) Plot of the Voronoi diagram for four samples. The color coding of the various sections represents the area of the section. Brighter areas are smaller. There is a clear trend for sections of small (large) size to group with other small (large) sized sections.}
\label{fig:voronoi}
\end{center}
\end{figure}

Given a set of nodes in space (or the plane) the Voronoi diagram \cite{art:Aurenhammer} is a sectioning into areas around each node where each section contains all the points that are closest to the node in its interior. This partitions space (plane) into sections filling the whole space (plane). The Delaunay network is the network where each node is linked to all the other nodes that it shares a Voronoi section border with. A visualization of this is given in figure \ref{fig:voronoi}, where the Voronoi diagram for four of the samples are plotted. The coloring of a given Voronoi section reflects the size of the section. Smaller sections have a brighter shade and larger sections have a darker shade. It can be seen that the sections are grouped according to size, making regions of the whole diagram that contains mainly large or small sections. The distribution of the areas of the various Voronoi sections have been calculated and fitted with a log-normal function. As in the facet case each sample has been normalized by its maximum area. The distribution follows the same general shape as the facet distribution, and they can both be seen in figure \ref{fig:area_facets}.

\begin{figure}
\centering
\begin{tabular}{l}
a) 
\\
\begin{minipage}{0.95\columnwidth}
\centering
\includegraphics[width=0.95\columnwidth, angle=0]{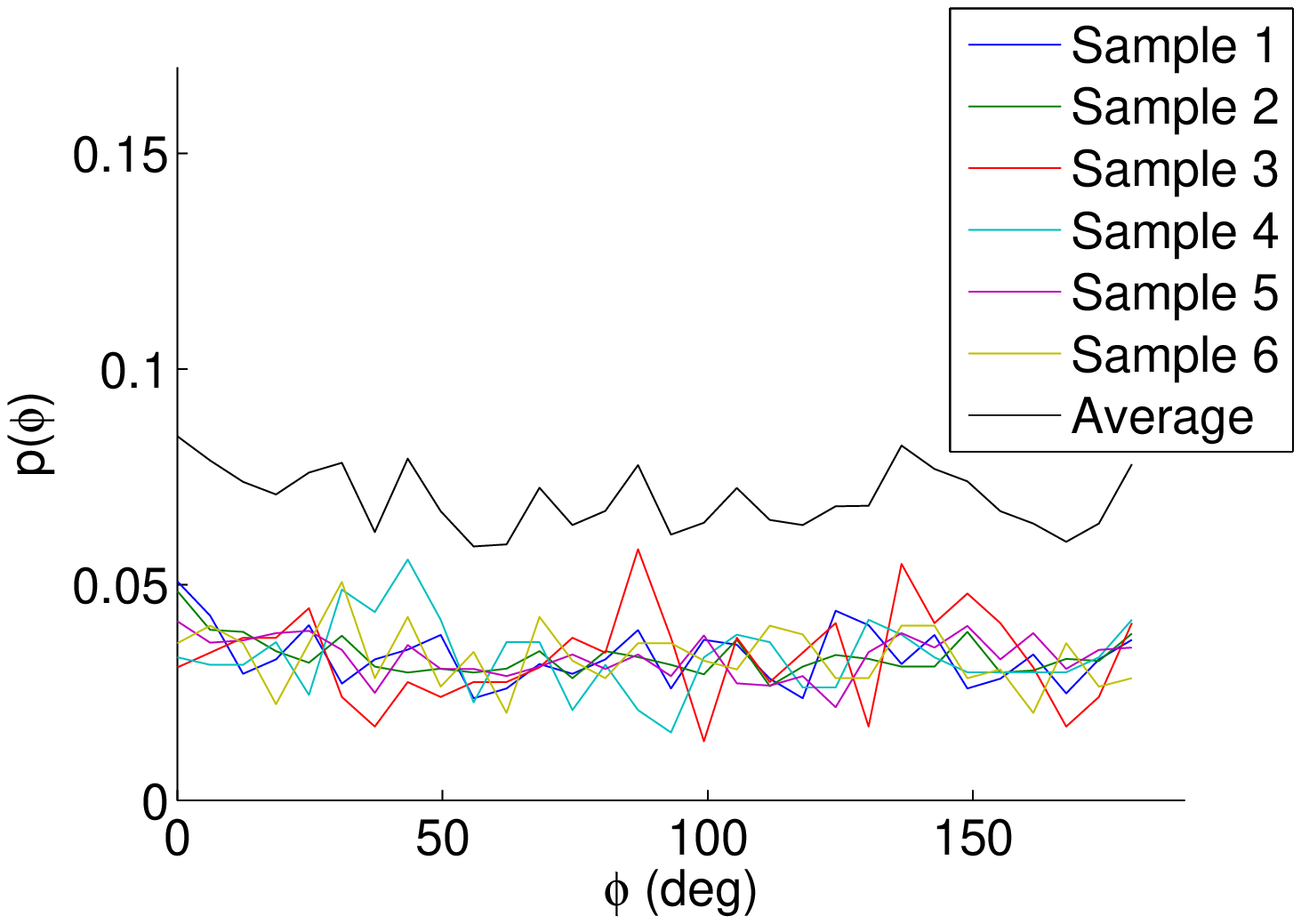} 
\end{minipage}
\\
b) 
\\
\begin{minipage}{0.95\columnwidth}
\centering
\includegraphics[width=0.95\columnwidth, angle=0]{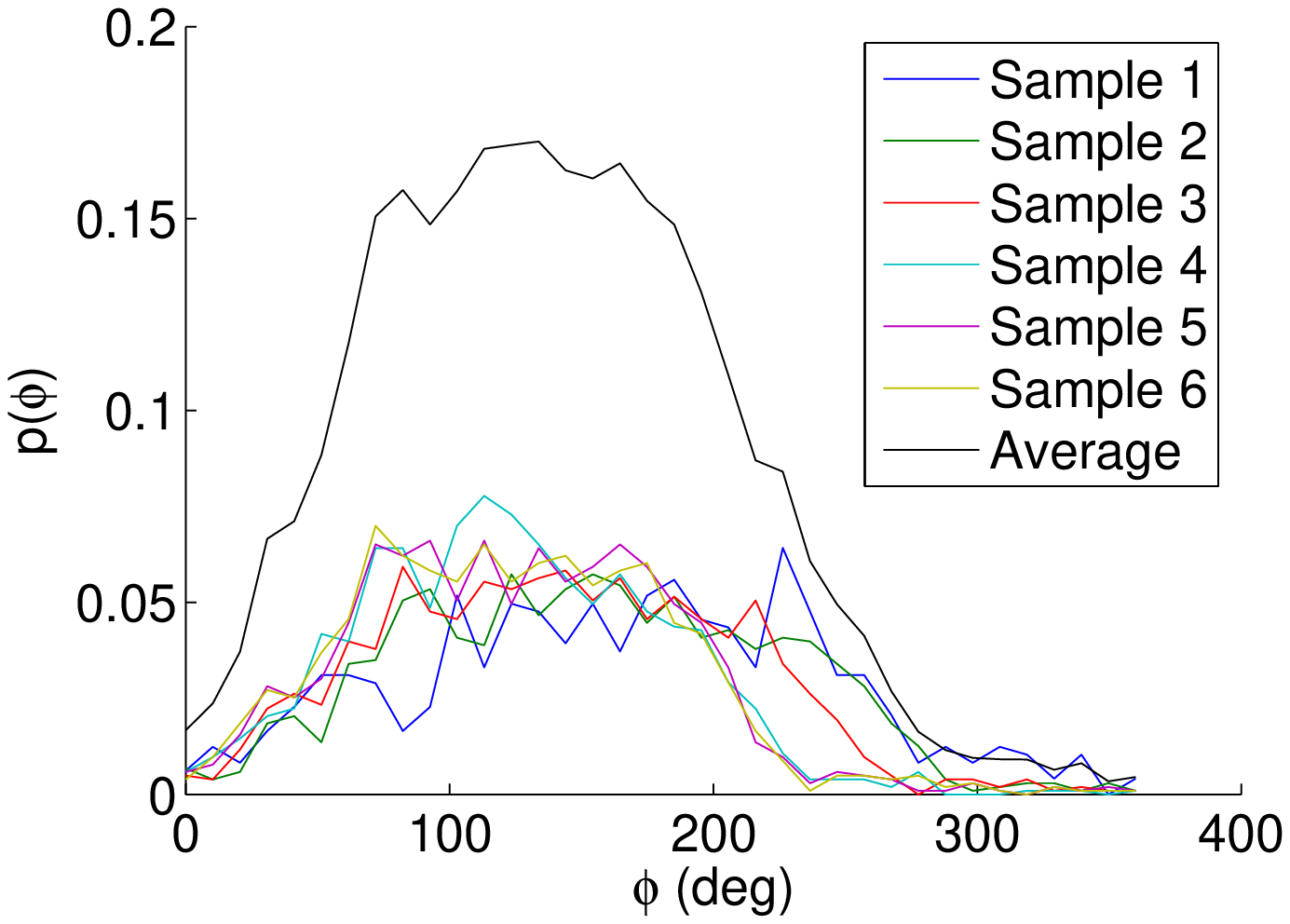}
\end{minipage}
\\
\end{tabular}
\caption{(Color On-line) a) Plot of the distribution of ridge angles, $p(\phi)$, as a function of angle $\phi$ for all the samples and the average (elevated for clarity). No trend is visible in the plot. b) Distribution of separation angles, $p(\phi)$, as a function of opening angel $\phi$ in 3 ridge nodes. The average is elevated for clarity.} 
\label{fig:collage1}
\end{figure}

\subsection{Angular distribution}

The angular distribution of the ridges relative to the border of the sample has been studied in order to detect any preferred ridge direction or ordering among the ridges with regard to direction. No such preferred direction or ordering was found, and the distribution of ridge angles was reasonably uniform, both for each sample and for the average. A plot of the binned ridge angle distribution can be seen in figure \ref{fig:collage1}. 

The distribution of angles between ridges in a three ridge cone (a node where three ridges meet, and hence form a cone like structure) have earlier been investigated both analytically and experimentally \cite{art:Blair,art:Cerda,art:Chaieb}. It has been reported that there is indications of preferred opening angles for such cones in the regions about $20\,^{\circ}$, $60\,^{\circ}$ and $110\,^{\circ}$, although all acquired distributions have been broad. All $k=3$ nodes have been investigated and the distribution of the ridge separation angles shows a broad distribution of angles with a maximum in the range between $100\,^{\circ}$ and $150\,^{\circ}$. There are no significant peaks in the distribution and this indicates a random ordering. However 32 \% of all the angles lies in the interval between $90\,^{\circ}$ and $150\,^{\circ}$. This suggests that the ridges tend to span out trying to separate themselves from each other. Recall that $120\,^{\circ}$ is the angle at which they are evenly separated. A plot of the distribution for all the samples and their average can be seen in figure \ref{fig:collage1}.

\section{Roughness}

\begin{figure}
\centering
\begin{tabular}{l}
a)
\\
\begin{minipage}{0.95\columnwidth}
\centering
\includegraphics[width=\columnwidth, angle=0]{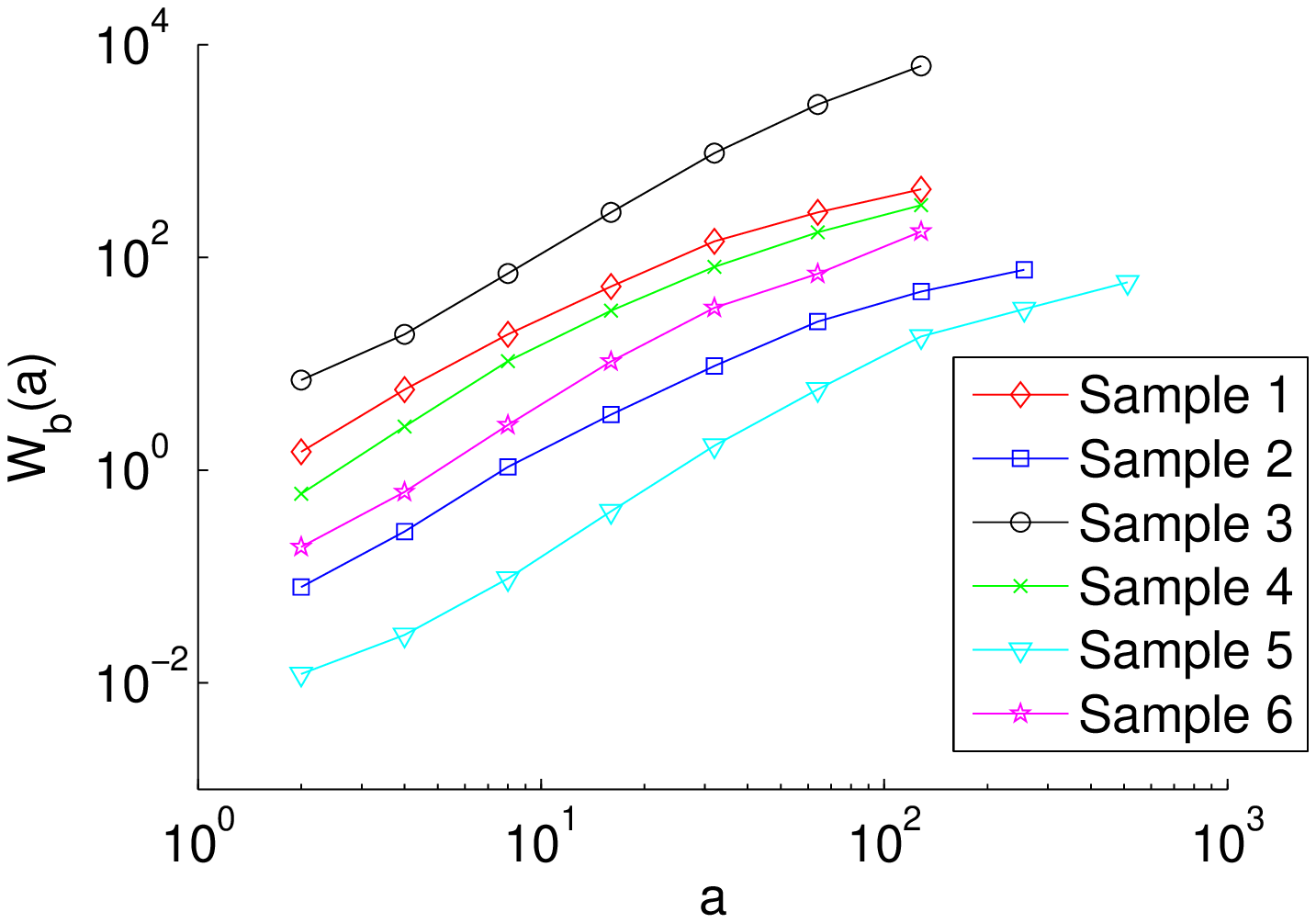}
\end{minipage}
\\
b)
\\
\begin{minipage}{0.95\columnwidth}
\centering
\includegraphics[width=\columnwidth, angle=0]{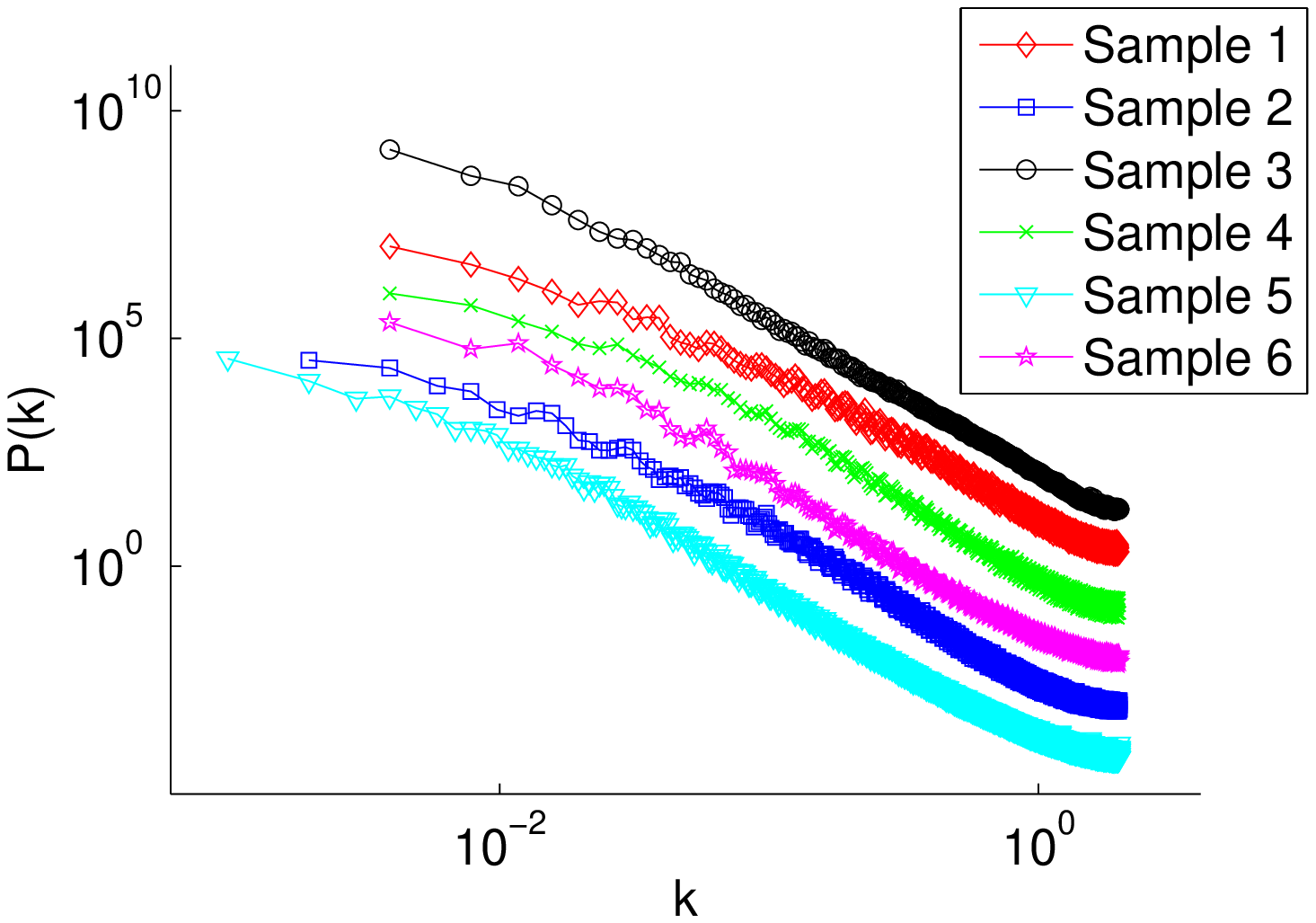}
\end{minipage}
\\
\end{tabular}
\caption{(Color On-line) a) Results from the AWC analysis showing the average wavelet coefficients, $W_b(a)$, as a function of scale $a$, the scale represents measured points. b) Results from the PSD analysis showing the power spectrum density, $P(k)$, as a function of spatial frequency $k$. In both a) and b) the data sets are vertically shifted for clarity.} 
\label{fig:collage2}
\end{figure}

The roughness of crumpled paper surfaces has been investigated before \cite{art:Blair,art:Plouraboue,art:Balankin}. These investigations have reported self-affine behavior, this means that the surface is statistically characterized by 

\begin{equation}
h(x) = \lambda^{-H}h(\lambda x),  
\end{equation}  

where $h(x)$ is the height of the profile at position $x$, $\lambda$ is a rescaling factor and $H$ is the Hurst exponent. We have investigated the one-dimensional profiles produced by the profilometer using the average wavelet coefficient (AWC) method \cite{art:Simonsen}, the power spectrum density (PSD) method \cite{art:Mehrabi} and the bridge method \cite{art:engoy}. The results from all the methods indicates that the crumpled paper forms a self-affine surface. Earlier works have reported a small scale region with a Hurst exponent $\textrm{H}_S \sim$ 1.0 and a large scale region with $\textrm{H}_L \sim$ 0.7 \cite{art:Blair,art:Balankin} and $\textrm{H}_L \sim$ 0.8 \cite{art:Plouraboue}. Our results follows the same trend in that there is a cross over scale between two scaling regimes. However we found the small scale exponent to be $H_{S} = 1.25 \pm 0.05$, indicating that the surface is asymptotically non-flat at these scales. Unfortunately the data did not give a robust value for $\textrm{H}_L$ because the sample size was too small compared to the crossover scale. The data did however indicate that $\textrm{H}_L < 1.0$ and in the range reported above. In figure \ref{fig:collage2}, results from the PSD and AWC methods can be seen.

\section{Conclusion}

The main points reported above are that the tail of the ridge length distribution is found to be well reproduced by a power-law distribution, and that the short ridges follows a log-normal distribution as reported earlier. The degree distribution has been shown not to have a power-law tail, but rather an exponential decay, and the networks have been found to be disassortative. The facet area distribution, the corresponding Voronoi diagram area distribution and the Delaunay vertex distribution have all been found to fit log-normal distributions.

\begin{acknowledgments}
This work have been supported by VISTA, a research cooperation between \textit{the Norwegian Academy of Science and Letters} and \textit{Statoil}. We would also like to thank J. \O. Bakke and R. Toussaint for fruitful discussions, and the referees for valuable input.
\end{acknowledgments}

\end{document}